\documentstyle[12pt]{article}
\begin{document}
\title{Non-Dopplerian cosmological redshift parameters
in a model of graviton-dusty universe}

\author
{Michael A. Ivanov\\
Physics Dept.,\\
Belarus State University of Informatics and Radioelectronics,\\
6 P. Brovka Street,  BY 220027, Minsk, Belarus.\\
E-mail: ivanovma@gw.bsuir.unibel.by}


\maketitle

\begin{abstract}
Possible effects are considered which would be caused by a
hypothetical super-strong interaction of photons or massive bodies
with single gravitons of the graviton background. If full
cosmological redshift magnitudes are caused by the interaction,
then the luminosity distance in a flat non-expanding universe as a
function of redshift is very similar to the specific function
which fits supernova cosmology data by Riess et al. From another
side, in this case every massive body, slowly moving relatively to
the background, would experience a constant acceleration,
proportional to the Hubble constant, of the same order
as a small additional acceleration of Pioneer 10, 11. \\
\end{abstract}

\section[1]{Introduction }
In this paper, possible manifestations of the graviton background
in a case of hypothetical super-strong gravitational quantum
interaction are considered. From one side, the author brings the
reasons that a quantum interaction of photons with the graviton
background would lead to redshifts of remote objects too. The
author considers a hypothesis about  an existence of the graviton
background to be independent from the standard cosmological model.
One cannot affirm that such an interaction is the only cause of
redshifts. It is possible, that the one gives a small contribution
to an effect magnitude only.  But we cannot exclude that such an
interaction with the graviton background would be enough to
explain the effect without an attraction of the big bang
hypothesis. Comparing the own model predictions with supernova
cosmology data by Riess et al \cite{mivanov-12a}, the author finds
here good accordance between the redshift model and observations.
\par
From another side, it is shown here, that every massive body, with a non-zero
velocity $v$ relatively to the isotropic graviton background,
should experience
a constant acceleration. If one assumes that a full observable redshift
magnitude is caused by such a quantum interaction with single gravitons,
then this acceleration will have
the same order of magnitude as a small additional acceleration of NASA
deep-space probes (Pioneer 10/11, Galileo, and Ulysses), about which it was
reported by Anderson's team \cite{mivanov-15}.
\par
For more details, one can find a preprint of my full original paper
\cite{mivanov-4}
in the e-print archive \cite{mivanov-3}.
\section[2]{An interaction of photons and of massive bodies
with the graviton background }

If the isotropic graviton background exists, then, due to photon
scattering on gravitons, average energy losses of a photon with an
energy  $E $ on a way $dr $ will be equal to $ dE=-aE dr,$ where
$a$ is a constant \cite{mivanov-5,mivanov-3,mivanov-4}. It is
shown here, that such an interaction with single gravitons should
be super-strong to provide a full redshift magnitude. As it was
reported by Anderson's team \cite{mivanov-15}, NASA deep-space
probes experience a small additional constant acceleration,
directed towards the Sun. It follows from an universality of
gravitational interaction, that not only photons, but all other
objects, moving relatively to the background, should loss their
energy too due to such a quantum interaction with gravitons. If
$a=H/c,$ massive bodies or particles must experience a
deceleration $w$ of the same order as an additional acceleration
of cosmic probes: $$w = - ac^{2}(1-v^{2}/c^{2}).$$ The
acceleration $w$ is directed against a body velocity in a special
system of reference, in which the graviton background is
isotropic. It is for small velocities: $$w \simeq - Hc \simeq -
4.8 \cdot 10^{-10} m/s^{2},$$ if Hubble's constant $H = 1.6 \cdot
10^{-18} s^{-1},$ that corresponds approximately to one half of
the observed acceleration for NASA probes.
\par
It is possible, that an annual periodic term in the residuals of the both
Pioneers (see plot B in the Figure 1 \cite{mivanov-15a}) may be caused by own
additional acceleration of the Earth under its motion relatively to the
graviton background.

\section[4a]{Comparison of the redshift model with supernova cosmology data}
In a case of flat non-expanding universe, a photon flux relaxation,
due to non-forehead collisions of photons with gravitons,
can be
characterized by a factor $b,$ so that the luminosity distance $D_{L}$ \cite
{mivanov-12a}
is equal in our model to: $$D_{L}=a^{-1} \ln(1+z)\cdot (1+z)^{(1+b)/2} \equiv a^{-1}f_{1}
(z;b),$$
where $z$ is a redshift. The theoretical estimation for $b$ is: $$b= 3/2+2/\pi
=2.137$$ \cite{mivanov-3,mivanov-4}. Thus, the redshift $$z=\exp(ar)-1$$
and the luminosity distance $D_{L}$
 are characterized in the model by two parameters: $H$ and $b$ ($r$ is a
geometrical distance). One can introduce an effective Hubble constant $$H_{eff}
\equiv c{dz}/{dr} = H(z+1);$$ in a language of expansion it can be interpreted
as "a current deceleration of the expansion".
\par
High-z Supernova Search Team data \cite{mivanov-12a} give us a possibility
to evaluate $H$
in our model. We can use one of the best fits of the function $D_{L}(z; H_{0},\Omega_
{M},\Omega_{\Lambda})$ from \cite{mivanov-12a} (see Eq.2 in
\cite{mivanov-12a}) with $\Omega_{M}=-0.5$ and
$\Omega_{\Lambda}=0,$ which is unphysical in the original work. For $1-
\Omega_{M}>0$ and $1+\Omega_{M}z>0,$ the function $D_{L}(z; H_{0},\Omega_
{M},\Omega_{\Lambda})$ is equal to:
$$D_{L}=a^{-1}(1+z) m^{-1} \sinh (\ln \vert {(k-m)}/{(k+m)} \vert-\ln
\vert {(1-m)}/{(1+m)}\vert)$$ $$\equiv a^{-1}f_{2}(z;
H_{0},\Omega_{M},\Omega_ {\Lambda}),$$ where $m \equiv
(1-\Omega_{M})^{1/2}, k \equiv (1+\Omega_{M}z) ^{1/2}.$ Assuming
$b=2.137,$ we can find $H$ from the connection: $$HD_{L}
/H_{0}D_{L}=f_{1}(z;b)/f_{2}(z;
H_{0},\Omega_{M},\Omega_{\Lambda})$$ (see Table). We see that
$H/H_{0} \simeq const,$ a deviation from an average value $<H>
\simeq 1.09H_{0}$ is less than $5\%.$
\begin{table}[h]
\begin{tabular}{ccccccccccc} \hline
$z$        &0.1  &0.2  &0.3  &0.4  &0.5  &0.6  &0.7  &0.8  &0.9
&1.0\\ \hline
$f_{1}$&0.110&0.242&0.396&0.570&0.765&0.983&1.222&1.480&1.759&2.058\\
\hline
$f_{2}$&0.103&0.219&0,359&0.511&0.677&0.863&1.074&1.301&1.565&1.854\\
\hline
$H/H_{0}$  &1.068&1.105&1.103&1.115&1.130&1.139&1.138&1.138&1.124&1.110\\
\hline
\end{tabular}
\end{table}
If one would suggest that $f_{1}(z;b)$ describes results of
observations in an expanding universe, one could conclude that it
is "an accelerating one". But a true conclusion may be strange:
our universe is not expanding, and redshifts have the
non-Dopplerian nature.

\section[7]{Conclusion }

If further investigations display that an anomalous NASA probes'
acceleration cannot be explained by some technical causes, left out of
account today, it will give a big push to a further development of
physics of particles. Both supernova cosmology data and the Anderson's team
discovery may change a gravity position in a hierarchy of known interactions,
and, possibly, give us a new chance to unify their description.
\par


\begin{thebibliography}{References}
\bibitem{mivanov-4}
Ivanov, M.A., {\it General Relativity and Gravitation} (the paper is accepted
by the editor and is scheduled for the March, 2001 issue of the journal).
\end{thebibliography}
\end{document}